\documentclass[iop]{emulateapj}
\usepackage{amssymb,amsmath}
\usepackage{color,hyperref}
\usepackage{amsmath}    
\usepackage{natbib}
\usepackage{graphicx}   
\usepackage{verbatim}   
\usepackage{color,soul}

\usepackage{subfigure}  
\usepackage{hyperref}   
\newcommand{\degree}{\ensuremath{^\circ}}

\newcommand\geqsim{\lower.73ex\hbox{$\sim$}\llap{\raise.4ex\hbox{$>$}}$\,$}
\newcommand\leqsim{\lower.73ex\hbox{$\sim$}\llap{\raise.4ex\hbox{$<$}}$\,$}

\begin{document}

\title{Luminous Red Galaxies: Selection and classification by combining optical and infrared photometry }
\author{
Abhishek~Prakash\altaffilmark{1},
Timothy~C.~Licquia\altaffilmark{1},
Jeffrey~A.~Newman\altaffilmark{1},
Sandhya~M.~Rao\altaffilmark{1}
}

\altaffiltext{1}{
PITT PACC, Department of Physics and Astronomy, 
University of Pittsburgh, Pittsburgh, PA 15260, USA.
}

\email{abp15@pitt.edu}

\begin{abstract}
We describe a new method of combining optical and infrared
photometry to select Luminous Red Galaxies (LRGs) at redshifts $z > 0.6$. We explore this technique using a combination of optical
photometry from CFHTLS and HST, infrared photometry from the WISE satellite, and spectroscopic or photometric redshifts from the DEEP2 Galaxy Redshift
Survey or COSMOS. We present a variety of methods for testing the success of our selection, and present methods for optimization given a set of rest-frame color and redshift requirements. We have tested this selection in two different
regions of the sky, the COSMOS and Extended Groth Strip (EGS) fields,
to reduce
the effect of cosmic/sample variance. We have used these methods to
assemble large samples of LRGs for two different ancillary programs as
a part of the SDSS-III/ BOSS spectroscopic survey.  This technique is now being used to select $\sim$600,000 LRG targets for SDSS-IV/eBOSS, which
began observations in Fall 2014, and will be adapted for the proposed
DESI survey. We have found these methods can select high-redshift LRGs
efficiently with minimal stellar contamination; this is extremely
difficult to achieve with selections that rely on optical photometry
alone. 
\end{abstract}

\section{Introduction} \label{sec:intro}

Luminous Red Galaxies (LRGs) are relatively massive ($\sim 10^{11-12} M_{\odot}$), generally elliptical systems comprised primarily of old stars; they are the most massive and luminous ($\geq 3 L^{\star}$) galaxies in the $z$ $
\leq $ 1 universe. These galaxies are expected to reside in massive dark matter halos, and thus cluster very strongly \citep[e.g.,][]{Pad2007}. Their strong clustering enhances the Baryon Acoustic Oscillation (BAO) signal. 
Combined with their intrinsic brightness, this makes them excellent probes of the large-scale-structure (LSS) of the Universe and a vital tool for cosmology. \\
\indent LRGs exhibit a strong 4000 \AA\ break in their spectral energy distributions \citep[SEDs;][]{Eisenstein2005}. At lower redshifts $({z} \leq 0.6)$, LRGs can be efficiently selected and their redshifts estimated using optical 
photometry alone, by taking advantage of this feature. This method was used to select LRG targets for the Sloan Digital Sky Survey (SDSS) and the SDSS-III / Baryon Oscillation Spectroscopic Survey (BOSS), as well as the 2dF-
SDSS LRG and QSO survey \citep[2SLAQ;][]{Eisenstein2001,Cannon2006}. However, this method becomes extremely difficult at greater distances as cosmic expansion redshifts the 4000 \AA\  break into longer wavelength filters, 
therefore requiring long imaging exposure times to overcome the brightness of the night sky in the near-infrared (NIR). A new method is required in order to efficiently select LRGs at higher redshifts. \\
\indent Old stellar populations exhibit global maxima in their SEDs at a rest-frame wavelength of 1.6 $\mu$m, corresponding to the minimum in the opacity of H- ions in their stellar atmospheres \citep{john1988}. Since the light 
measured from LRGs is predominantly produced by old stars, we expect this feature to dominate their overall SEDs. This enables efficient selection of LRGs at higher redshifts.\\
\indent In this paper, we demonstrate that a simple cut in optical-infrared color-color space provides an efficient method for differentiating LRGs from other types of objects. The methods described here will be applied for selecting 
LRG targets for next generation spectroscopic surveys like The Extended Baryon Oscillation Spectroscopic Survey (eBOSS) and the Dark Energy Spectroscopic Instrument (DESI) Survey, and may be easily adapted to meet the 
needs of future prospects.\\ 
 \indent This paper is organized as follows: In Section 2, we describe the construction of samples used to test LRG selection methods, including both the imaging and spectroscopic datasets used. In Section 3, we present a simple 
selection method for identifying LRGs based on optical and infrared photometry and analyze the efficiency of this selection method for a set of nominal selection cuts applied to our sample. In Section 4, we explore methods for 
optimizing the LRG selection algorithm by adjusting the parameters of our cuts in color-color space. In Section 5, we summarize our results and conclude with plans for future work. For this work, we assume a standard $\Lambda
$CDM cosmology with $H_{0}$=100\textit{h} km s$^{-1}$ Mpc$^{-1}$, $\Omega_M = 0.3$, and $\Omega_\Lambda = 0.7$.

\section{Data} \label{sec:data}
In this paper, we make use of 5 cross-matched catalogs that cover two different regions of the sky, the Extended Groth Strip (EGS), with 214.0\degree $<$ $\alpha$ $<$ 215.70\degree and 52.14\degree $<$ $\delta$ $<$ 53.22\degree, and the COSMOS field, with 
149.41\degree $<$ $\alpha$ $<$ 150.82\degree and 1.49\degree $<$ $\delta$ $<$ 2.92\degree. These two regions have been surveyed by a variety of telescopes, providing photometry over a wide range of wavelengths. We have cross-matched objects based on 
their positions on the sky as recorded by each survey. In order to avoid duplicate matches, we match each object in the catalog with the lowest surface density to its nearest neighbors in the denser catalogs that are closer 
than 1.5 arcseconds. For DEEP 2 objects in the EGS, we use the cross-identifications provided by \cite{Dan2013}.  Below, we briefly describe each of the catalogs used in this study.
\subsection{Optical Photometry}
\textbf{Canada-France-Hawaii Telescope Legacy Survey (CFHT LS)}: CFHT LS consist of two parts. The Wide Survey covered $\sim150$ deg$^{2}$ divided over 4 fields with magnitude limits (50\% completeness for point 
sources) of $ u^{\star}\sim 26.0, g^{\prime} \sim 26.5, r^{\prime}\sim 25.9, i^{\prime} \sim 25.7$, and  $z^{\prime} \sim 24.6 $. The Deep Survey consists of 4 fields of 1 deg$^{2}$ area each, each with magnitude limits of $ u^{\star} 
\sim 27.5, g^{\prime} \sim 27.9, r^{\prime} \sim 27.7, i^{\prime}\sim 27.4$, and $z^{\prime}\sim 26.2 $. We use both the D2 Deep field which lies within the COSMOS region, and D3, which overlaps with EGS. We also use both the Wide survey and the Deep survey in the EGS.\\
\indent We use the COSMOS \textit{ugriz} magnitudes and their corresponding errors from the CFHT LS catalogs produced by \citet{Gwyn2011}, which were created using the \textit{MegaPipe} data pipeline at the Canadian 
Astronomy Data Centre.\\
\indent CFHT LS uses MegaCam filters which are slightly redder than their SDSS counterparts. We convert CFHT LS photometry to SDSS pass-bands by inverting the filter relations given in Equations 1-5. The relations for the 
\textit{g}, \textit{r}, \textit{i} and \textit{z}-bands come from analyses by the SuperNova Legacy Survey (SNLS) group.\footnote{http://www.astro.uvic.ca/~pritchet/SN/Calib/ColourTerms-2006Jun19/index.html} The relation for 
the \textit{u} band is taken from the CFHT web pages.\footnote{http://cfht.hawaii.edu/Instruments/Imaging/MegaPrime \\ /generalinformation.html} \footnote{http://www2.cadc-ccda.hia-iha.nrc-cnrc.gc.ca/en/megapipe/ \\ docs/filt.html} The transformed equations are:

\begin{align}
\textit{u}_{SDSS}&= \textit{u}_{Mega} + 0.181 (\textit{u}_{Mega} - \textit{g}_{Mega}), \\
\textit{g}_{SDSS}&= \textit{g}_{Mega} + 0.195 (\textit{g}_{Mega} - \textit{r}_{Mega}),\\
\textit{r}_{SDSS}&= \textit{r}_{Mega}  + 0.011 (\textit{g}_{Mega} - \textit{r}_{Mega}), \\
\textit{i}_{SDSS}&= \textit{i}_{Mega} + 0.001 (\textit{r}_{Mega} - \textit{i}_{Mega}), \rm{and}\\ 
\textit{z}_{SDSS}&= \textit{z}_{Mega} + 0.099 (\textit{i}_{Mega} - \textit{z}_{Mega}),
\end{align}
where \textit{u}$_{Mega}$, \textit{g}$_{Mega}$, \textit{r}$_{Mega}$, \textit{i}$_{Mega}$, and \textit{z}$_{Mega}$ represent the \textit{ugriz} magnitudes measured by CFHT LS and \textit{u}$_{SDSS}$, \textit{g}$_{SDSS}$, \textit{r}$_{SDSS}$, \textit{i}$_{SDSS}$, and \textit{z}$_{SDSS}$ are the standard SDSS magnitudes. The resulting SDSS-passband magnitudes are then corrected for Galactic extinction using the dust map of \citet{Sfd1998}, hereafter 
SFD. To calculate the extinction in a given band, A($\lambda$), we interpolate the standard total-to-selective extinction ratios, i.e.  A($\lambda$)/E(B-V) from Table 6 of \citet{Sfd1998} for the effective wavelengths given in the filter 
list of CFHT LS.\footnote{http://www2.cadc-ccda.hia-iha.nrc-cnrc.gc.ca/en/megapipe/ \\ docs/filt.html} We obtain E(B-V) values from the SFD dust map \citep{Sfd1998} via the routine $dust\_getval.pro$ provided in the idlutils 
package.\footnote{http://www.sdss3.org/dr8/software/idlutils.php}

\subsection{Infrared photometry} 
\textbf{Wide-Field Infrared Survey Explorer (WISE) catalog}: \textit{WISE} completed a mid-infrared survey of the entire sky by July 2010 in four infrared channels, labeled \textit{W1,W2,W3} and \textit{W4}, centered at 3.4, 4.6, 12, 
and 22 $\mu$m, respectively. This was achieved using a 40 cm telescope with much higher sensitivity than previous infrared survey missions. WISE achieved $5\sigma$ point source sensitivities better than 0.08, 0.11, 1, and 6 
mJy, corresponding to 19.1423, 18.7966, 16.4001, and 14.4547 AB magnitudes,\footnote{M$_{AB}$ = - 2.5  $\times$ $\log_{10}$(F$_{\nu}$/ 3631 Jy)} with angular resolutions of 6.1, 6.4, 6.5, and 12.0 arcseconds in the 
\textit{W1,W2,W3} and \textit{W4} channels, respectively \citep{wright2010}.\\
 \indent A detection by WISE is required for an object to be in our catalog. This restriction will have negligible effect on LRGs, since they are bright in the \textit{W1}-band but greatly reduces the number of objects to which we must 
apply our selection cuts. Based on color-magnitude diagram, we observe that all \textit{z} $<$ 20.5 LRGs are detected in \textit{W1} band at greater than 5-sigma. 3.4 micron (\textit{W1}) magnitudes are taken from the publicly available WISE All-Sky Data Release catalog of \cite{wright2010}. We convert these to the AB magnitude system and correct them for 
reddening using the SFD dust map \citep{Sfd1998} and interpolate extinction ratios, much as above.

\subsection{Redshifts}
\textbf{COSMOS}: The COSMOS photometric redshift (`photo-$z$') catalog from \citet{Ilbert2008} is a magnitude-limited catalog with \textit{I}$<25$. This catalog provides photometric redshifts over the $\sim 2$ deg$^{2}$ 
COSMOS field. The redshifts are computed using 30 bands covering the UV (GALEX), Visible-NIR (Subaru, CFHT, UKIRT) and mid-IR (Spitzer/IRAC). A $\chi^{2}$ template-fitting method yields photo-$z$ estimates which are 
calibrated with spectroscopic redshift measurements from VLT-VIMOS and Keck-DEIMOS. For details of photo-$z$ determinations and accuracy, see \citet{Mobasher2007}.\\ 
\indent \textbf{EGS}: In the EGS we use the DEEP2 spectroscopic catalog. DEEP2 is a high-resolution redshift survey of $\sim$53,000 galaxies at redshifts $z \ge 0.7$ using the DEIMOS spectrograph at Keck Observatory 
\citep{jeff2013}. The survey covers an area of 2.8 deg$^{2}$ over four different fields. DEEP2 targeted galaxies 
brighter than \textit{R}$_{AB}$ $\sim$ 24.1 with a spectral resolution of \textit{R} ($=$ $\Delta\lambda/ \lambda$)$\sim$ 6000 and a central wavelength of 7800\AA\ \citep{jeff2013}.\\ 
 \indent Since the DEEP2 catalogs only provide BRI photometry, \cite{Dan2013} have created a catalog to supplement them with \textit{ugriz} photometry from CFHTLS and SDSS. Each catalog is cross-matched by position on the 
sky in order to assign \textit{ugriz} photometry to objects in the DEEP2 catalogs. We use the \cite{Dan2013} catalog in the EGS field to obtain \textit{ugriz} photometry. We correct this photometry for extinction as described above in Section 2.1.\\
\indent All objects in our datasets are required to have reliable redshifts. This is important as this information is used in determining the rest-frame colors of galaxies. For the COSMOS field, we use the photometric redshifts, 
\textit{zp\_gal}, taken directly from \citet{Ilbert2008}. We don't consider objects with \texttt{zp\_best}$= NULL$ as these are the objects in masked area. For the galaxies in the EGS, we make use of the heliocentric reference-frame spectroscopic redshift, \textit{ZHELIO}, provided in the DEEP2 extended photometry catalog of 
\cite{Dan2013}. We also ensure that each galaxy in our sample has a securely measured redshift ( i.e., we require redshift quality flags, \textit{ZQUALITY} of 3 or 4 in DEEP2).

\subsection{Object type identification} 
\textbf{COSMOS}: In order to distinguish stars, galaxies, and X-ray sources within the COSMOS field from each other, we apply a variety of cuts based upon the photometric redshift estimates,  \textit{zp\_{best}}, as well as the 
reduced chi-squared value associated with the separate star and galaxy template fits to each object's SED, \textit{Chi\_{star}} and \textit{Chi\_{gal}}.\footnote{Variables defined the same way they appear in the catalogs} The parameters we use have been provided by \citet{Ilbert2008}. To identify 
different objects, we use the criteria:\\
\begin{flushleft}
\begin{eqnarray}
\rm{\textbf{X-ray\ sources}}: \textit{zp\_{best}} > 9,\\
\nonumber \\
\rm{\textbf{Galaxies}}: (\textit{Chi\_{gal}}  < \textit{Chi\_{star}}) \nonumber \\
\ AND\ (0.011 < \textit{zp\_{best}}  < 9),\\
\nonumber \\
\rm{\textbf{Stars}}: (\textit{chi\_{gal}}  > \textit{chi\_{star}}) \nonumber \\
 \ OR\ ( \textit{chi\_{gal}}  < \textit{chi\_{star}}) \nonumber \\
 \ AND\  (\textit{zp\_{best}}  <  0.011 \ OR\ \textit{zp\_{best}}  > 9))
\end{eqnarray}
\end{flushleft}
These criteria distinguish galaxies from stars and X-ray sources. An object is flagged as a star if its SED is best fit with a stellar template or if it yields an extremely low redshift in case where a galaxy template is the better fit. An 
object identified as a galaxy should not only fit the galaxy template best but also yield a redshift between 0.011 and 9. Once the objects are identified, \textit{zp\_{final}} is our redshift indicator and its value is set to the photometric redshift, 
\textit{zp\_{gal}}, if the object is identified as a galaxy. The \textit{zp\_{final}} value is set to $0.0$ for stars and $-9.99$ for X-ray sources.\\
\indent \textbf{EGS}: We use the Hubble Space Telescope-Advanced Camera for Surveys (HST-ACS) general catalog for objects in the EGS field. The HST-ACS General Catalog is a photometric and morphological catalog created using 
publicly available data obtained with the ACS instrument on the Hubble Space Telescope (HST). This provides a large sample of objects with reliable structure measurements. It includes approximately 470,000 sources originally 
observed in a variety of sky surveys, including the All-Wavelength Extended Groth Strip International Survey (AEGIS), COSMOS, GEMS, and GOODS \citep{Griffith2012}. A single Sersic model for each object is assumed for 
deriving quantitative structural parameters (e.g., surface brightness and effective radius).\\
\indent Our goal is to be able to estimate the stellar contamination in the EGS. DEEP2 avoided targeting stars, so we can not assess this from the spectroscopic sample on its own. No such effort is required for the COSMOS field 
since the catalog of \cite{Ilbert2008} contains both stars and galaxies. We use the same definition as \citet{Griffith2012} for identifying compact objects (presumed to be stars) based upon their larger surface brightness, $mu\_{HI}$, 
and lower effective half-light radius, $RE\_{Galfit}\_{HI}$ (for more details, see Figure 5 of \cite{Griffith2012}).
Specifically, we identify objects with mu\_{HI} $<$ 18.5 or (mu\_{HI} $>$ 18.5 \& RE\_{Galfit}\_{HI} $<$ 0.03) as stars, where RE\_{Galfit}\_{HI} is given in arc-seconds.
Once the number of the stars is determined, we assume that the fraction of objects which are stars is uniform over the entire EGS. This gives us the total number of selected objects by our color-cut (galaxies, x-ray sources, and stars) which is then used to calculate the normalized Figure of Merit, see Figure~\ref{fig:com_fom_v}. This step is necessary since the HST-ACS general catalog of \citet{Griffith2012} covers only a portion of the entire EGS.\\

\section{Method of LRG selection}
\label{sec:method}
\begin{figure}[h]
\centering
\includegraphics[scale=0.39]{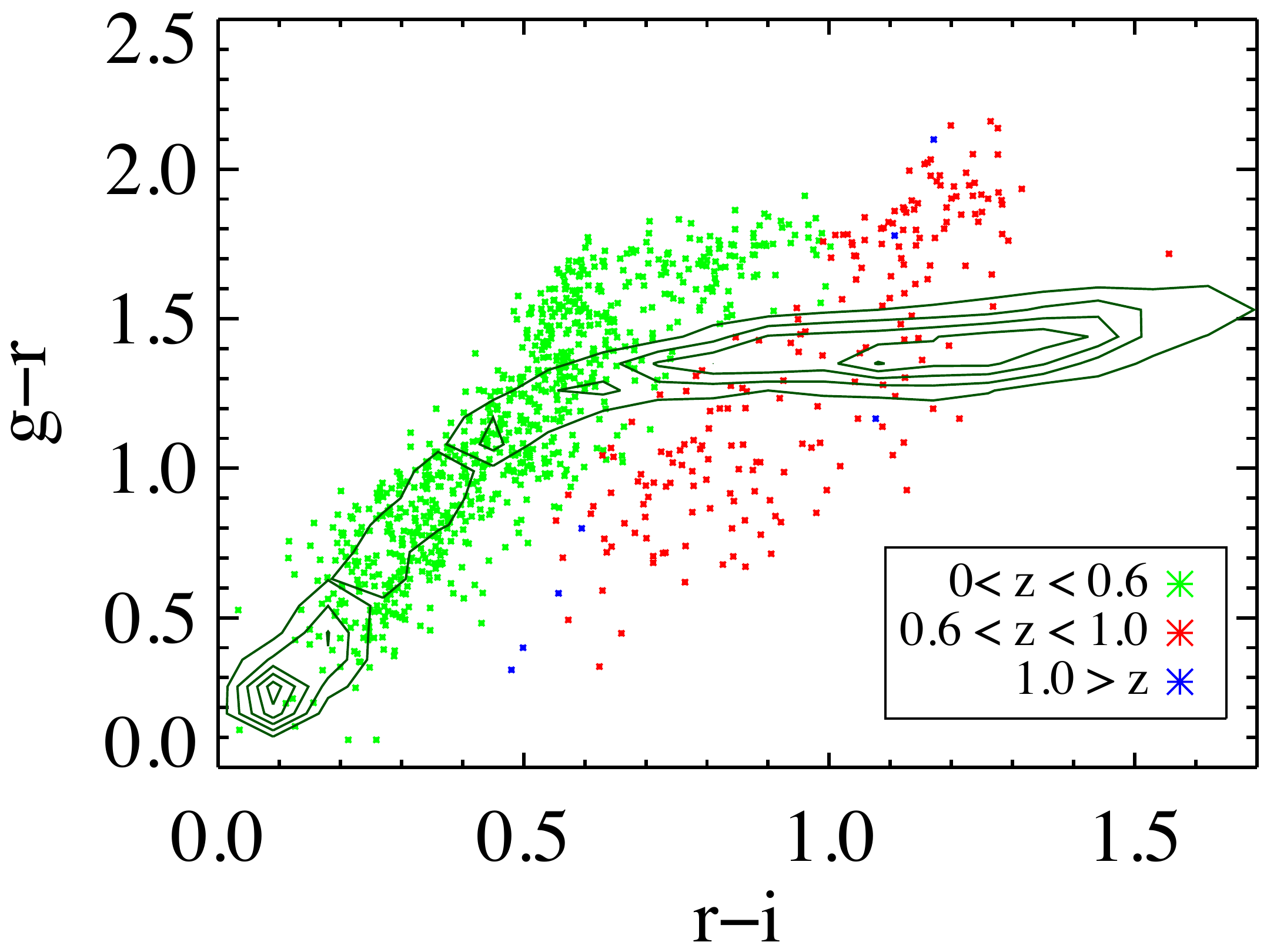}
\caption  {\textsl{\textit{g-r} vs \textit{r-i} optical color-color plot of galaxies (data points) observed by CFHT LS with COSMOS photometric redshifts. Star density contours are overplotted to show the overlap with galaxies. This overlap makes LRG selection difficult at higher redshifts, requiring a new method of selecting them.}}
\label{fig:gri_contour}
\end{figure}

\begin{figure}[h]
\centering
	\includegraphics[scale=0.39]{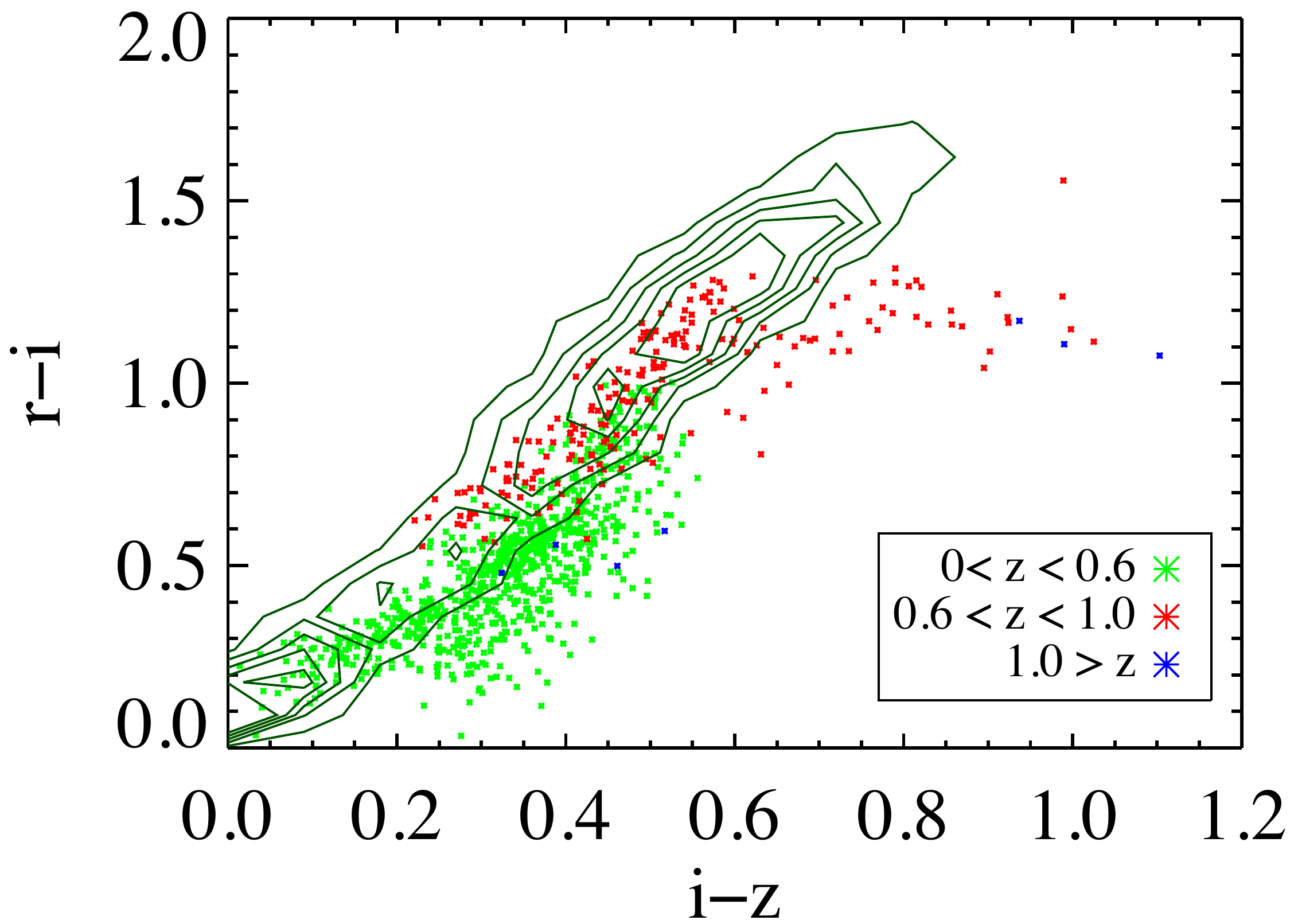}
	\caption  {\textsl{\textit{r-i} vs \textit{i-z} optical color-color plot of galaxies (data points) observed by CFHT LS and the COSMOS survey, similar to Figure 2. Star density contours are overplotted to show the overlap with 
galaxies. Stars overlap almost entirely with high-redshift LRGs.}} 
\label{fig:riz_contour}
\end{figure}

Our goal is to develop a method for selecting LRGs at high redshift, i.e., $z > 0.6$. One of the main challenges in LRG selection based on optical photometry alone is stellar contamination. The color overlap of stars with galaxies is 
illustrated in Figures~\ref{fig:gri_contour} and~\ref{fig:riz_contour}, where contours depicting the density of stars are overlaid on the locations of galaxies (shown as dots) in \textit{g-r} vs. \textit{r-i} (Figure~\ref{fig:gri_contour}) and  
\textit{r-i} vs. \textit{i-z} (Figure~\ref{fig:riz_contour}) color-color plots. The strong overlap of these populations makes them difficult to separate cleanly. 

\begin{figure}[h]
\centering
\label{fig:sdss_wise_image}
\includegraphics[scale=0.29]{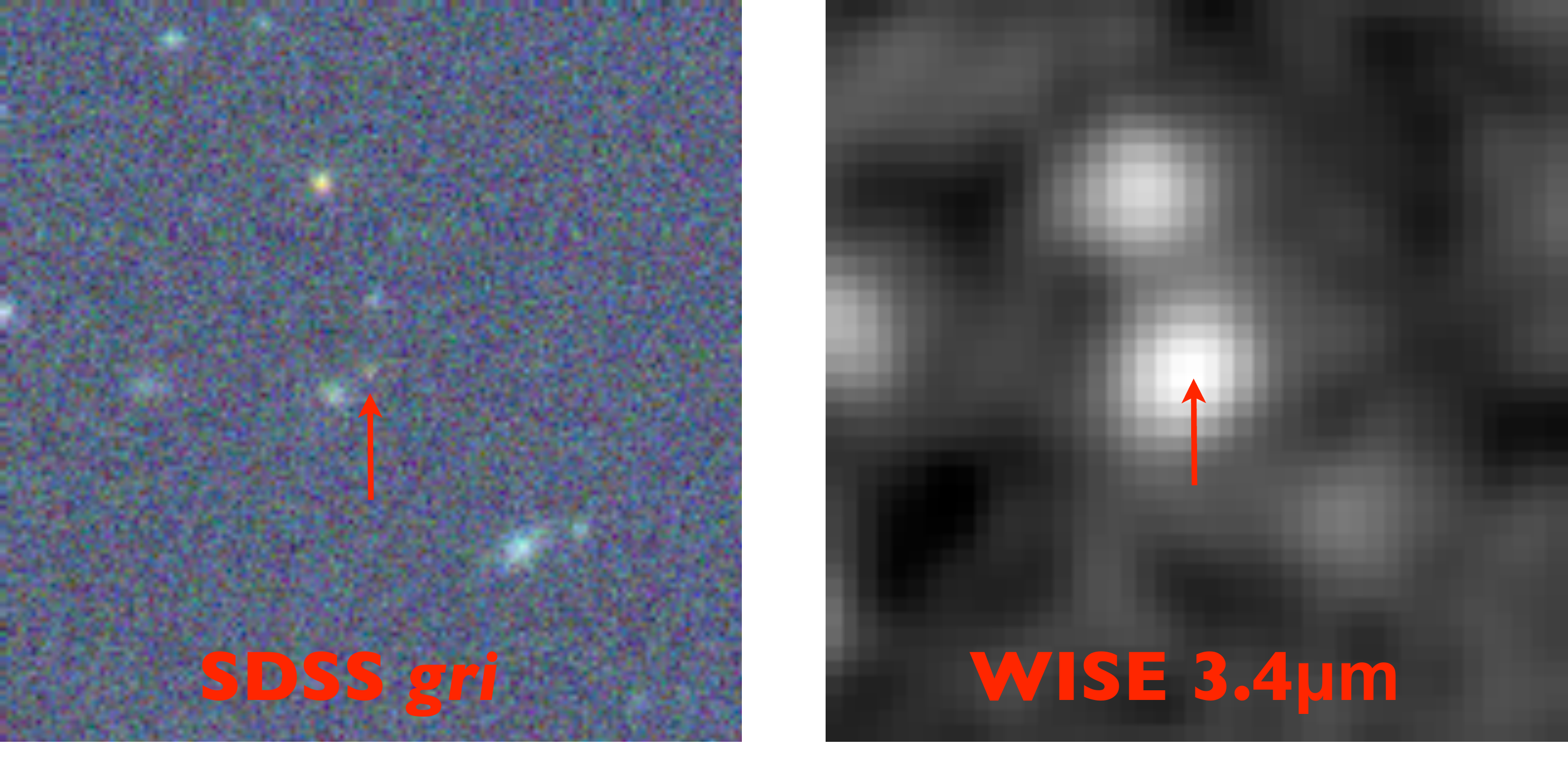}
\caption  {\textsl{1 arcminute square SDSS and WISE (3.4 micron) images of a $z \sim 1$ LRG. This object is at the 5 sigma detection limit in SDSS \textit{r} and \textit{i} but dominates the WISE image. Due to the redshifted `1.6 
$\mu$m bump', LRGs at $z \sim 1$ look much brighter in the WISE 3.4 micron band than at optical wavelengths, providing a new method of selecting them.}}
\end{figure}

\begin{figure}[hbtp]
\centering
	\includegraphics[scale=0.39]{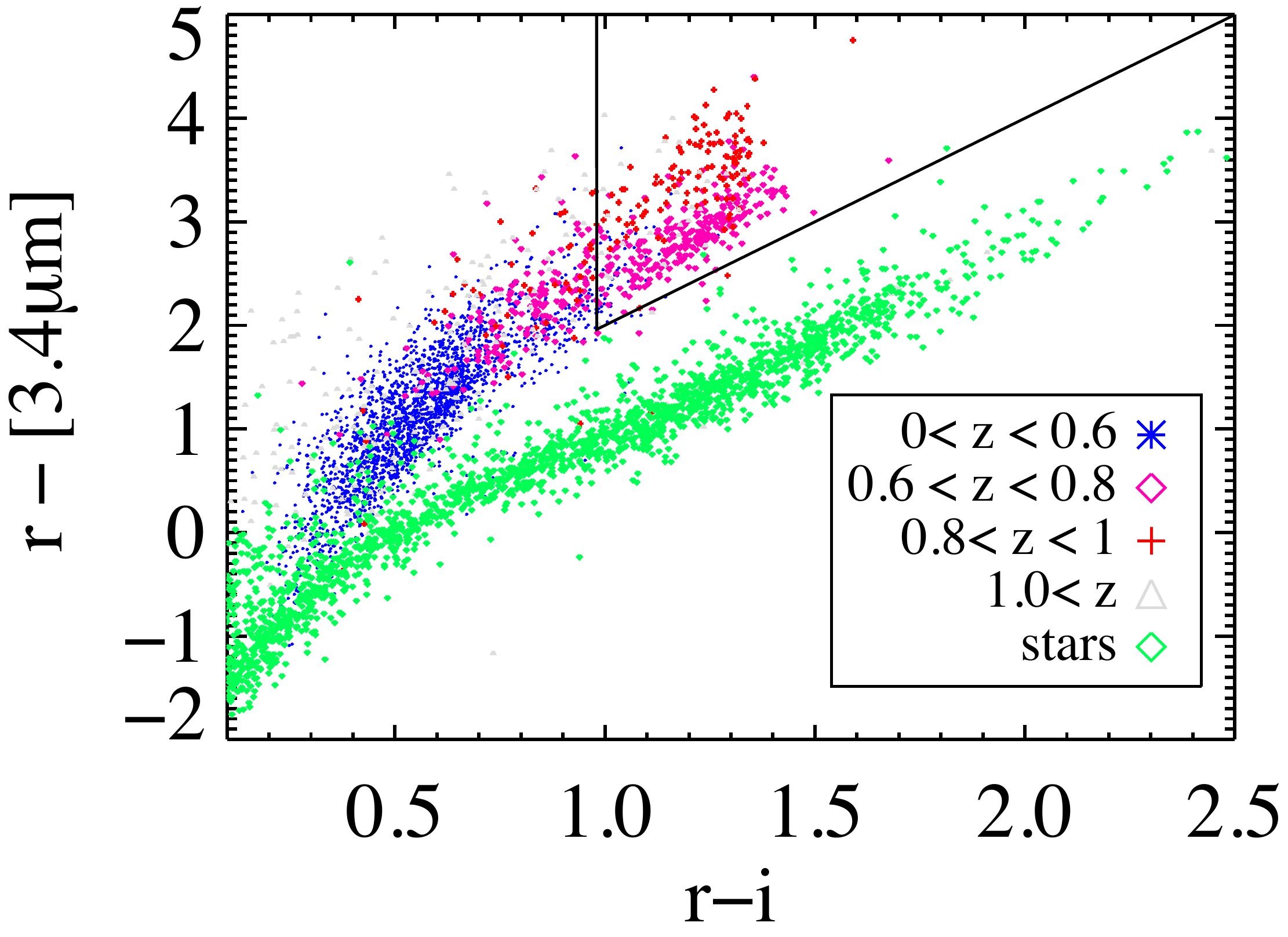}
	\caption{ \textsl {Optical/Infrared color-color plot for galaxies observed by WISE, CFHT LS and the COSMOS survey. Blue symbols represent all the galaxies at z $<$ 0.6. Pink diamonds represent galaxies at z $>$ 0.6 while red crosses represent z $>$ 0.8. Cyan triangles represent galaxies at higher redshift of z $>$ 1.0. Stars are represented by green diamonds. The triangular region represents the LRG selection region used in Figures~\ref{fig:ubz_cos}, \ref{fig:zhist_cos} and \ref{fig:roc_cos}. }}
\label{fig:rw1_cos}
\end{figure}

In this paper, we present a new technique for identifying high-redshift LRGs which combines optical and infrared photometry. The lowest wavelength channel of imaging from the WISE satellite is centered around 3.4 $\mu$m. This overlaps with the red-shifted `1.6 $\mu$m bump' at redshifts of $z \sim 0.6-1$, causing LRGs at those redshifts to appear very bright in this band. This phenomenon is illustrated in 
Figure~\ref{fig:sdss_wise_image}, where a typical LRG at $z \sim 1.0 $, which is barely detected in a $1\arcmin$ $\times$ $1\arcmin$ region of SDSS optical imaging, is the brightest object in the WISE NIR image of the same area. The relative brightness of  LRGs in the WISE 3.4 $\mu$m band compared to the optical bands increases monotonically up to $z \sim 1$ and then declines past $z \sim 1.1$ (at which point optically bright LRGs become rare).
\\
\indent To match the expected spectroscopic depth of DESI LRGs, we restrict the dataset for all analyses in this paper to those objects which have SDSS \textit{z}-band magnitude \textit{z} $<$ 20.5. We now present a new technique which combines both optical and infrared photometry as a means of selecting galaxies that are intrinsically red and at high redshift while circumventing most stellar contamination. In Figure~\ref{fig:rw1_cos}, we 
show both stars and galaxies in a plot of \textit{r-W1} color based on WISE and SDSS-passband photometry as a function of their SDSS \textit{r-i} color.  Here we can easily see that the two populations (stars, shown as green 
diamonds, and galaxies, shown as all other colored diamonds) exhibit a natural separation in the NIR-optical color space. In fact, the separation between the two populations grows as a function of galaxy redshift, allowing clean 
identification of the LRGs at higher redshifts, \textit{z} $>$ 0.6. Simultaneously, \textit{r-i} color increases with increasing redshift as the 4000 \AA\ break shifts redward, particularly for intrinsically red galaxies, allowing a selection 
specifically for intrinsically red, higher-redshift objects.\\

\begin{figure}[h]
\centering
	\includegraphics[scale=0.38]{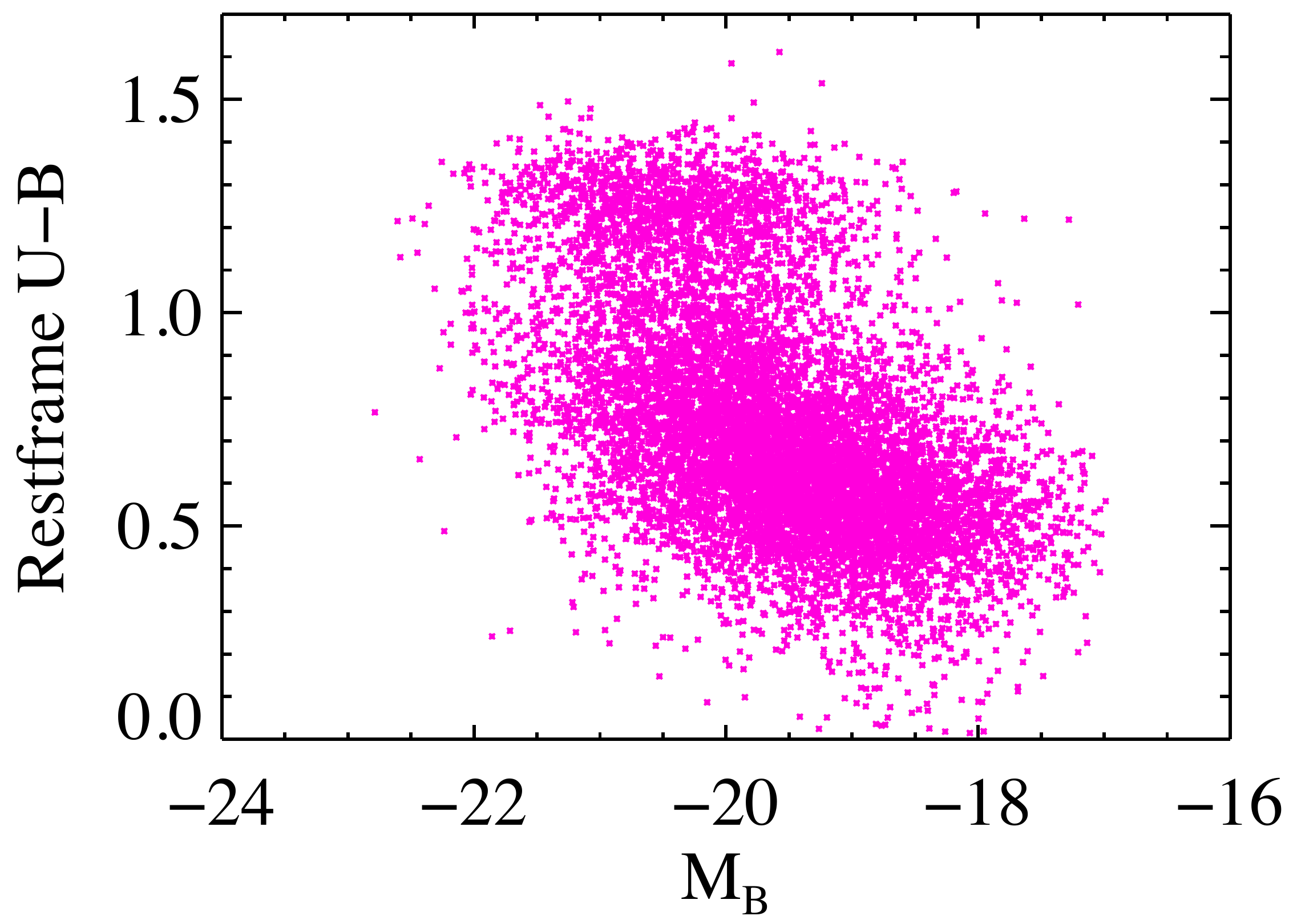}
	\caption{\textsl{Standard color-magnitude diagram of DEEP2 DR4 galaxies in redshift range $ 0.5 < z < 1.0$. Restframe U-B color is plotted on the y-axis, and the absolute B-band magnitude is plotted on the x-axis. We clearly see two different overdensities, one 
around  U-B $\sim$ 0.4 - 0.7  and another around U-B $\sim$ 1.25. The galaxies with U-B $\geq$ 1.0 are commonly referred to as the red-sequence galaxies while galaxies having U-B $<$ 1.0 form the blue cloud. The low density region in between these overdensities is generally referred to as the green valley. Restframe U-B colors and absolute magnitudes are obtained using the \texttt{k-correct} package which assumes $h = 1$; see \citet{blanton-kc-2007} for details. Choosing a different value of $h$ will not change U-B, just M$_{B}$.}}
\label{fig:color_mag}
\end{figure}

\begin{figure}[h]
\centering
	\includegraphics[scale=0.38]{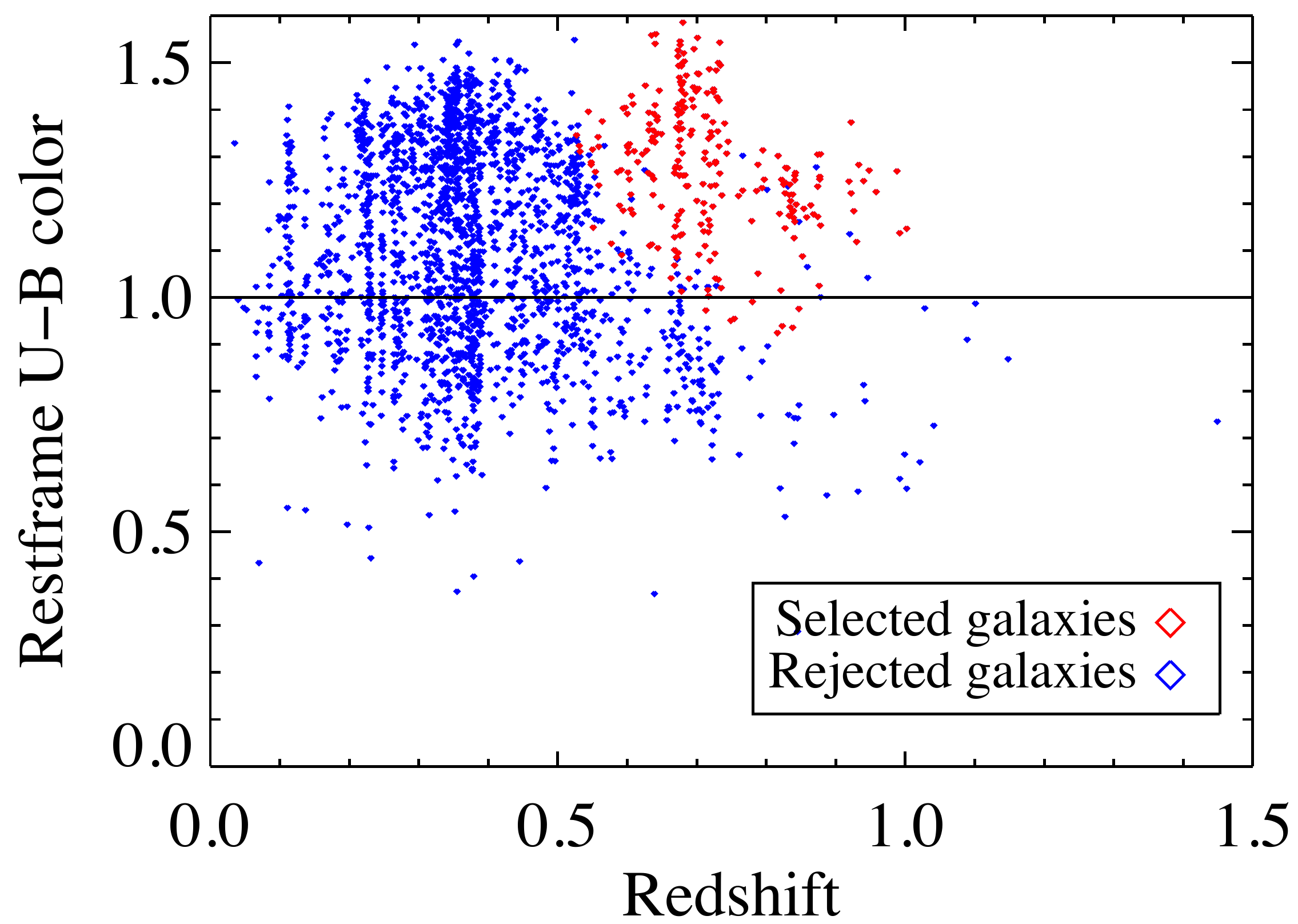}
	\caption{\textsl{Rest-frame U-B colors of WISE-detected, z $< $ 20.5 galaxies as a function of redshift. Red dots represent the galaxies from the LRG selection region of Figure~\ref{fig:rw1_cos}). Blue dots represent the 
galaxies excluded by this color selection. The red sequence corresponds well to those galaxies with U-B $\geq 1.0$. Of the galaxies selected by our color cut, 85.8\% have U-B greater than 1.0; most selected objects that miss this cut are only slightly bluer than the red sequence.}}
\label{fig:ubz_cos}
\end{figure}

\begin{figure}[t]
\centering
	\includegraphics[scale=0.38]{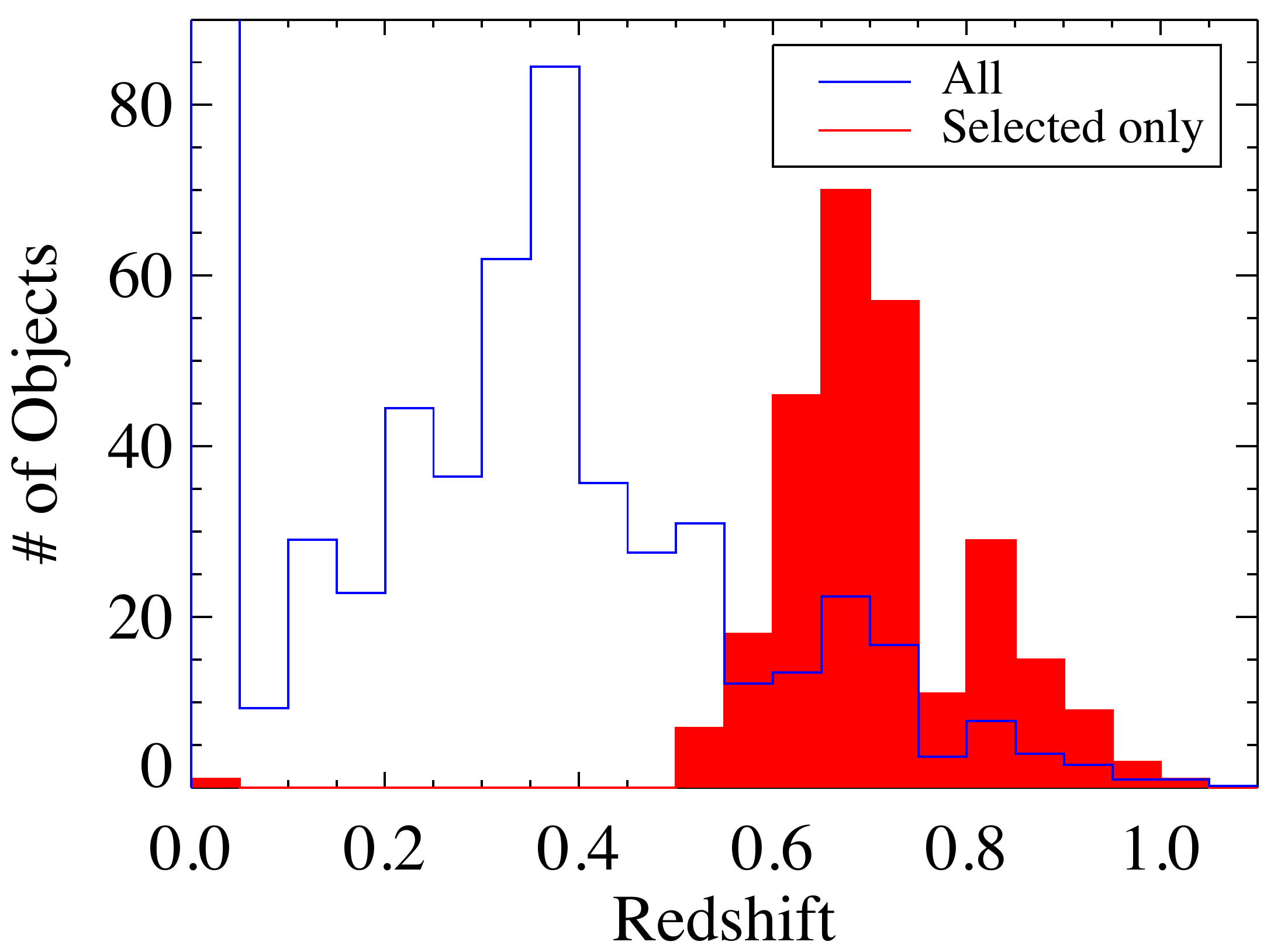}
	\caption{\textsl{Redshift histogram of WISE-detected, z $<$ 20.5 galaxies. The blue histogram depicts the redshift distribution of the full WISE-detected, z $<$ 20.5 sample, with an arbitrary renormalization applied. The red histogram represents galaxies from the LRG selection region of Figure~\ref{fig:rw1_cos}. The selected objects predominantly fall in the redshift range 0.6 $<$ $z$ $<$ 1, with a modest 0.63\% contamination by stars (appearing at $z \sim 0 $).}}
\label{fig:zhist_cos}
\end{figure}

As a result, a simple cut in the optical-infrared color-color plot enables us to efficiently select LRGs at higher redshifts, rejecting bluer galaxies, lower-redshift objects, and stars. As a nominal scenario, we select all objects that have 
both \textit{r-i} $> 0.98$ and \textit{r-W1} $> 2.0$ $\times$ \textit{(r-i)}, where \textit{r} and \textit{i} are extinction-corrected SDSS magnitudes and \textit{W1} is the magnitude in the WISE 3.4 micron pass-band on the AB system 
(Figure~\ref{fig:rw1_cos}). We have determined these cuts through visual optimization by examining the populations in Figure~\ref{fig:rw1_cos}.\\
 \indent Overall, our LRG color-cut selection has three free parameters: the minimum allowed \textit{r-i} color (corresponding to the vertical line in Figure~\ref{fig:rw1_cos}), and the slope and intercept of the line determining the 
minimum allowed \textit{r-W1} color at a given \textit{r-i} color (corresponding to the inclined line in Figure~\ref{fig:rw1_cos}). The latter of these two criteria determines the degree to which stars are rejected from our sample of 
LRGs. The \textit{r-i} cut will mostly affect the properties of the galaxies we select (e.g., their redshift distribution). We investigate the performance of this color-cut in Figure~\ref{fig:ubz_cos},~\ref{fig:zhist_cos}, and~\ref{fig:roc_cos}. For an object to be classified as a high-redshift LRG, we generally require it to have both a rest-frame color $U-B$ $\geq$ 1 and a redshift $z  > $ 0.6, though we also consider other redshift thresholds. We 
use the \texttt{k-correct} package to obtain rest-frame $U-B$ color for all galaxies; see \citet{blanton-kc-2007} for details. \\
\indent To further justify our choice of redness threshold, $U-B$ $\geq$ 1, we have 
plotted the standard color-magnitude diagram for DEEP2 galaxies in Figure~\ref{fig:color_mag}, clearly showing the red-sequence, blue sequence and the green valley. From Figure~\ref{fig:ubz_cos}, we observe that with our nominal color-cut, 85.8\% of the galaxies selected have rest-frame $U-B$ $\geq$ 1, indicating that they are intrinsically red in rest-frame color, adopting the same 
criterion employed by \citet{Gerke2007}. The remaining galaxies selected are still relatively red and massive, but have some ongoing star formation.  

In Figure~\ref{fig:zhist_cos}, we show a histogram of the redshifts measured for each of the objects selected, indicating that 77.6\% of the galaxies selected by our nominal cuts fall within the redshift range of interest, i.e., $0.6 
\geq z \geq 1.0$, and $ < 1\%$ are stars.\\

\section{Optimization of LRGs selection} \label{sec:lrg_opt}
\subsection{Optimization using Receiver Operating Characteristic} \label{sec:roc_opt}
\begin{figure}[h]
\centering
\includegraphics[scale=0.39]{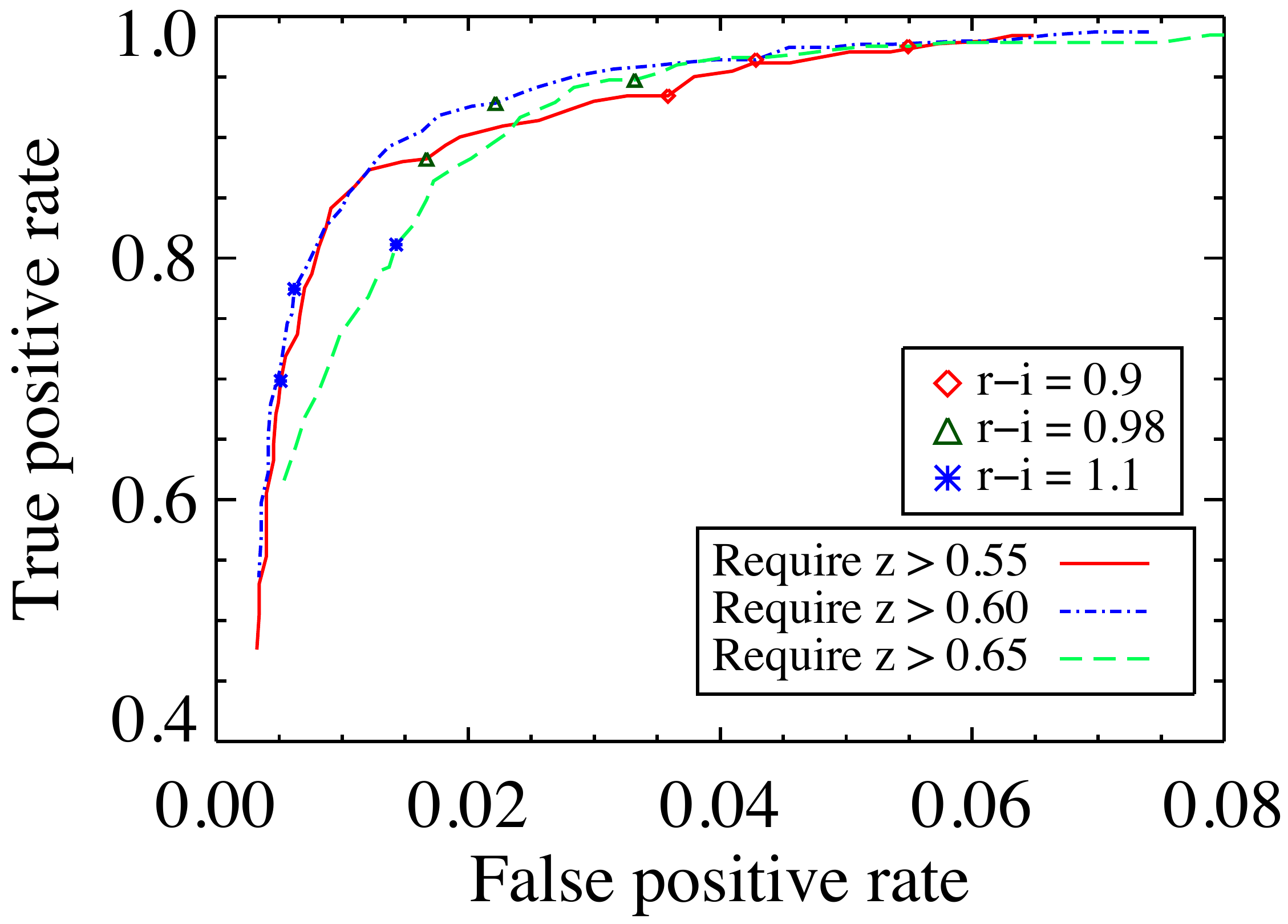}
\caption{\textsl{The Receiver Operating Characteristic plot for changes in the minimum allowed \textit{r-i} for selecting LRGs
over the range $ 0.8 < \textit{r-i} < 1.2 $. For each curve, we only consider intrinsically red objects that
have redshifts above a different specified minimum redshift, $z$, to be high-$z$ LRGs. The brown triangles represent the
performance at the threshold of $\textit{r-i} = 0.98 $. Similarly, the red diamonds and the blue stars represent the performance at the threshold of $\textit{r-i} = 0.9$ and $\textit{r-i} = 1.1$, respectively. Based on the commonly-used Area Under Curve criterion, our color cuts perform slightly better for LRGs with minimum redshift of $ z \sim 0.6 $ than 0.55 or 0.65.}}
\label{fig:roc_cos}
\end{figure}

\indent As a first attempt to optimize the efficiency of our selection method, we begin by varying the minimum allowed \textit{r-i} color (corresponding to shifting the vertical line in Figure~\ref{fig:rw1_cos}), while keeping all other parameters fixed. We interpret the results of varying this color criterion using a Receiver Operating Characteristic (ROC) plot, as shown in Figure~\ref{fig:roc_cos}. The ROC plot provides a visualization of the performance of a 
classification system; in our case, this quantifies our ability to segregate out the high-redshift LRGs in contrast to galaxies which are either blue or at lower redshift (`non-LRGs' for short). Each individual curve shows the result of 
using a different threshold on the minimum-allowed $z$ to be a `high-redshift' LRG; we only consider LRGs above the desired minimum redshift as our target population. \\
\indent The y-axis of the ROC plot represents the True Positive Rate (TPR), also known as the `sensitivity'. The TPR is defined as the fraction of all true high-redshift LRGs in the underlying sample that are within a given color cut. 
One of our main goals is to maximize the TPR. Of course, if the minimum-allowed \textit{r-i} color was shifted so blue as to select all galaxies this would be achieved by definition. However, there is a cost associated with doing this; 
we would select many galaxies that are not LRGs. This misidentification is quantified as the False Positive Rate (FPR) which is plotted on the \textit{x}-axis of the ROC plot. The FPR is the fraction of all non-LRGs (in our case, all 
\textit{z} $<20.5$, WISE-selected objects that are blue ($U-B$ $<$ 1) and/or below the desired redshift) that are placed in the LRG sample by a given color cut. Our goal is to minimize this quantity when varying the color cuts used 
to pick our sample while at the same time maximizing the TPR. \\
\indent One common way of assessing the performance of a selection algorithm is the Area Under Curve (AUC) diagnostic \citep{ROC1982}. This is calculated by integrating the ROC curve over all FPRs. Here, we assess the 
efficiency of selecting LRGs using 3 different redshift thresholds, $z \geq$ 0.55, 0.6, or 0.65. We have varied the minimum allowed \textit{r-i} color over the range in values from 0.8 to 1.2, and calculated the TPR and the FPR for 
each selection. Figure~\ref{fig:roc_cos} shows that, with our chosen cuts, we are able to attain a TPR of $85-95 \%$ (depending on the choice of the minimum allowed 'high' redshift, threshold $z$) while at the same time keeping 
the FPR below $\sim$$3\%$. Based on the AUC, we conclude that our selection algorithm performs best for a threshold redshift of $z \geq$ 0.6 (corresponding to the blue curve in Figure~\ref{fig:roc_cos}), as it encompasses the 
maximum area.
\subsection{Optimization using Figure Of Merit for large scale structure studies} \label{sec:fom_opt}
\begin{figure}[h]
\centering
\includegraphics[scale=0.39]{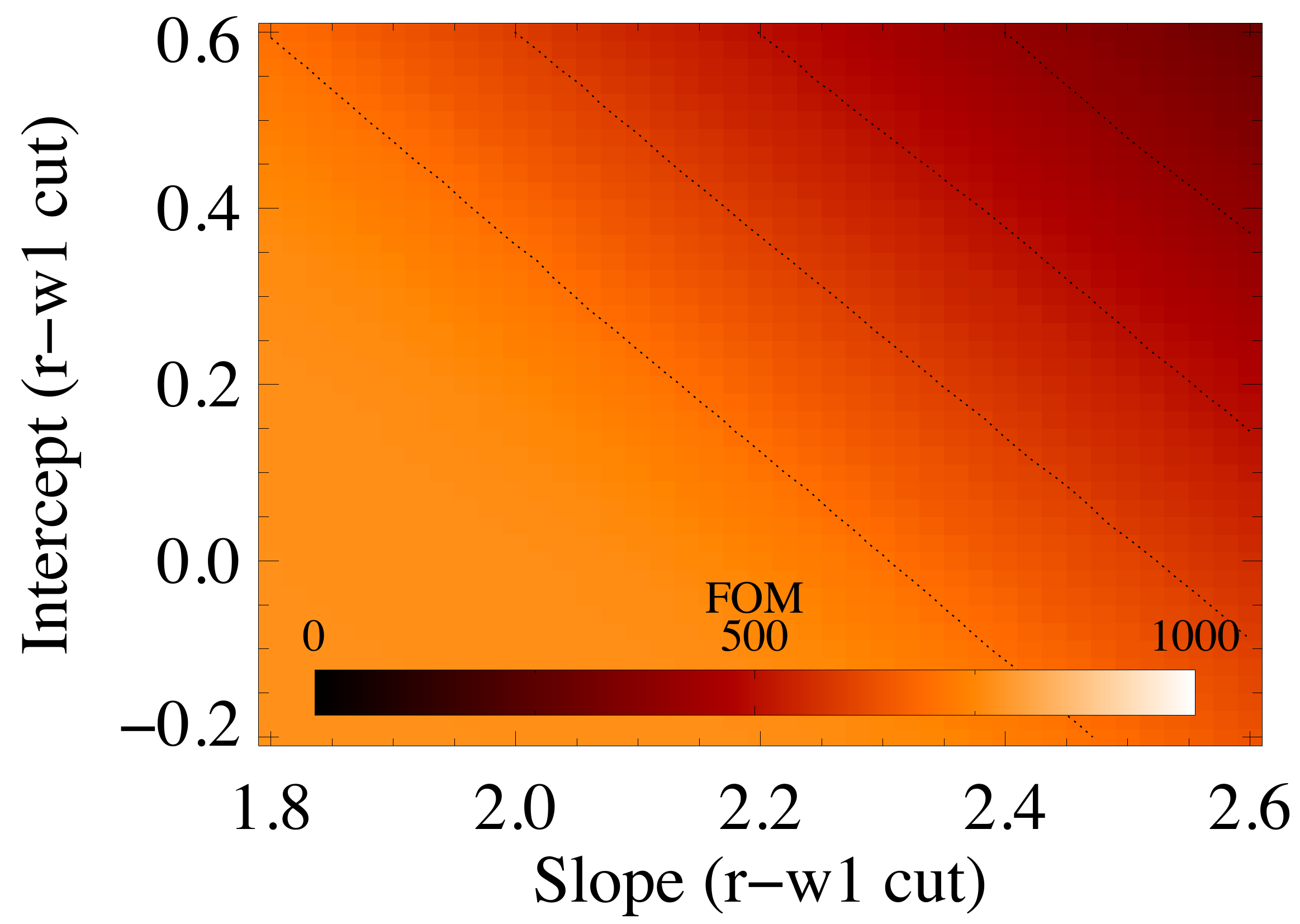}
\caption{\textsl{The LRG selection figure of merit (see Equation~\ref{eqn:fom_cos}) as a function of the slope and intercept of the r-W1 cut, using a fixed r-i threshold of 0.98. The visible pattern and the overplotted contours of constant FOM show that these parameters can be effectively replaced by a single linear combination of the two, as in Equation~\ref{eqn:slope_int_eqn}.} }
\label{fig:slope_int_cos}
\end{figure}

\begin{figure}[h]
\centering
\includegraphics[scale=0.38]{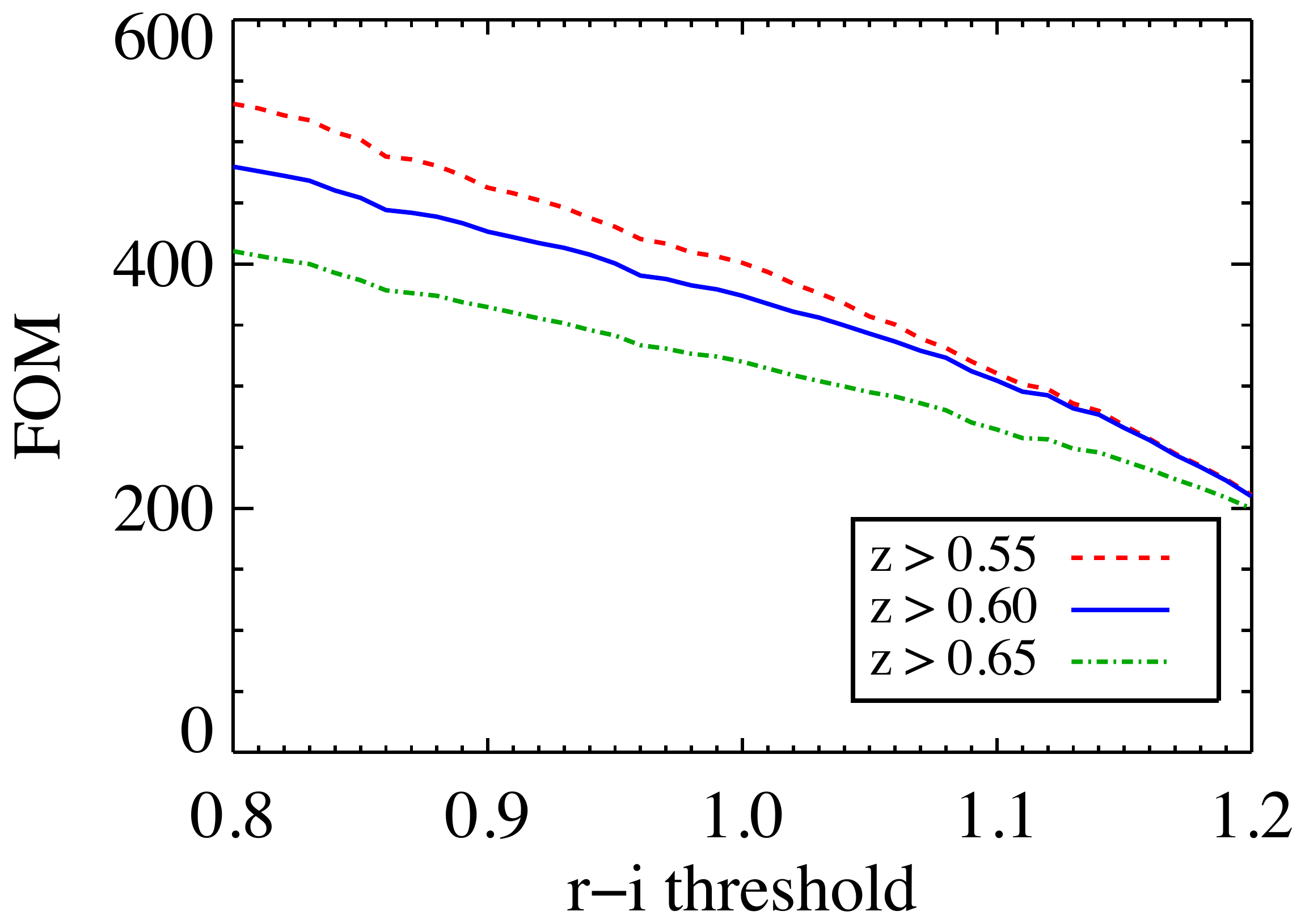}\\
\includegraphics[scale=0.38]{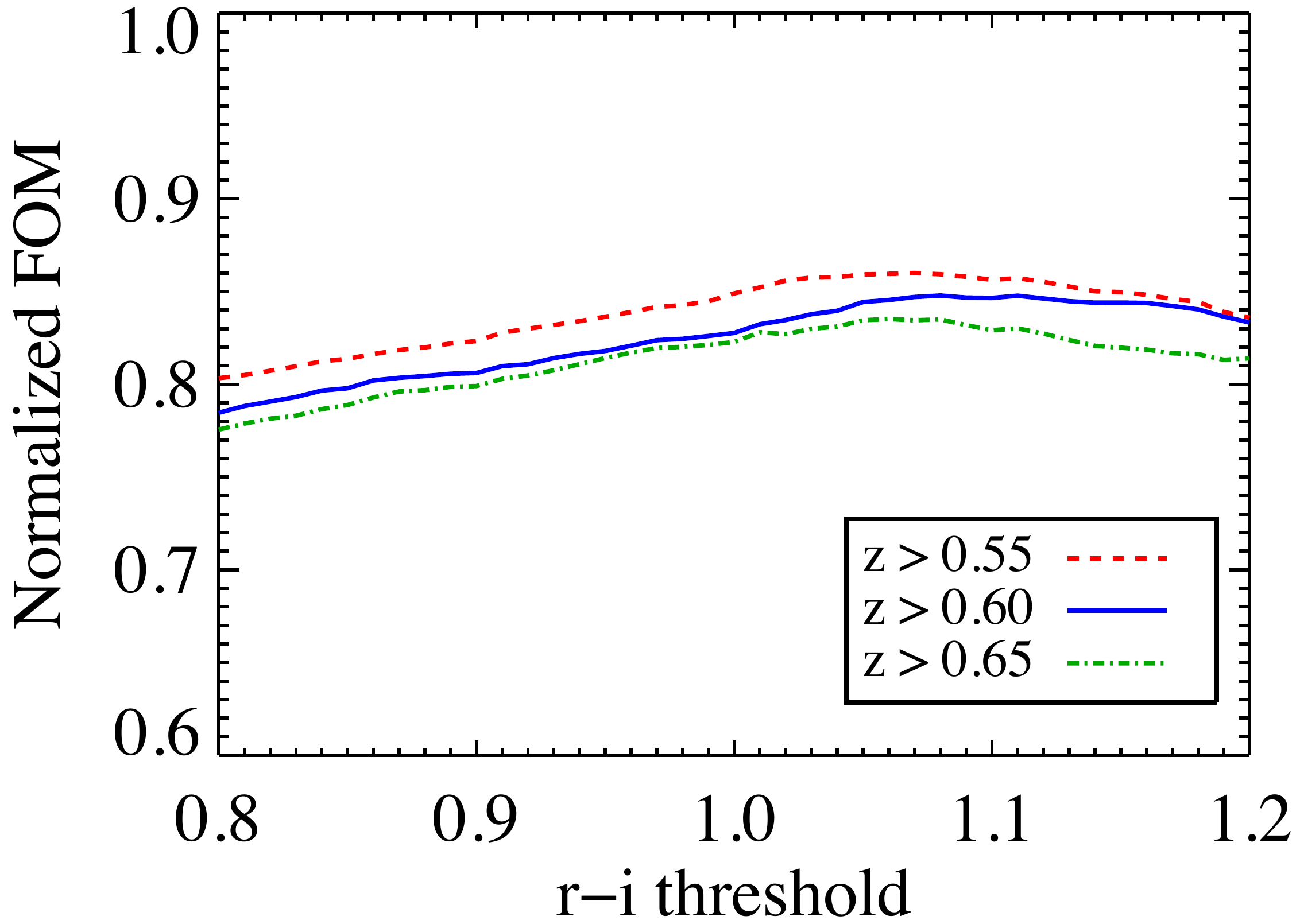}
\caption{\textsl{The maximum FOM for a given \textit{r-i} color threshold for z $<$ 20.5 galaxies, using data in the COSMOS region. Each individual curve shows the result of using a different choice for the 
minimum-allowed $z$ for a high redshift LRG. In the top panel, the monotonic decrease in FOM as the threshold \textit{r-i} color moves redward (corresponding to moving the vertical line in Figure~\ref{fig:rw1_cos} to the right) is consistent with 
our expectations, since we select a decreasing number of both LRGs and high-$z$ blue galaxies as the \textit{r-i} cut is moved to the right. In the bottom panel, we analyze the normalized FOM as the threshold \textit{r-i} color moves redward. For \textit{r-i} < 1.05, as the color cut moves redder, it selects a higher fraction of high-$z$ LRGs, increasing the purity of the sample and hence the normalized FoM monotonically.} }
\label{fig:fom_ri}
\end{figure}

To optimize the efficiency of our methods further, we attempt to account for the contribution to cosmological analyses from those non-LRG objects which are selected. While LRGs are the prime targets we are after, those galaxies 
which are not red and have a redshift of $z \geq 0.6$ (hereafter, `high-$z$ blue galaxies'), still provide useful information. To assess this, we define a Figure Of Merit (FOM) as\\
 \begin{eqnarray}   
     FOM = a \times n_{LRGs} + b \times n_{high-z\ blue\ gals}, 
\label{eqn:fom_cos}
\end{eqnarray}
where $a$ and $b$ are constants weighting the targets, and $n_{LRGs}$ and $n_{high-z\ blue\ gals}$ represent the number density (number per unit area) of LRGs and high-redshift blue galaxies, respectively, for a given set of color cut criteria. Since LRGs are our prime targets, we assign $a=1.0$ and $ b < a $. For the purpose of this paper, we adopt $b = 0.75$ as an example; a more ideal weighting would set $b/a$ according to the relative contribution of each class of objects towards, e.g., the uncertainty in the BAO scale. We assume that stars, X-ray sources (which tend to be at higher $z$), and galaxies at $z \leq$ 0.6 contribute nothing towards the FOM. We want to optimize not only the total FOM but also the FOM per object which we will refer to as the Normalized FOM. The total FOM represents the total constraining power of the whole sample and will increase even when we select lower-value high-$z$ blue galaxies. On the other hand, the Normalized FOM will increase only when LRGs make up a higher fraction of the sample and, therefore, can be more useful for optimization. Both should be considered; selecting all the objects regardless of color would yield a large total FOM but little value per object; while an extremely restrictive selection could have normalized FOM$\sim 1$, but have little total constraining power due to the small number of objects chosen. We then 
create a 3-dimensional grid to tabulate the FOM for each possible combination of the minimum allowed \textit{r-i} and the slope and intercept of the line marking the minimum allowed \textit{r-W1} for a given \textit{r-i} color. Figure~\ref{fig:roc_cos} shows that our nominal cut of $ \textit{r-i} = 0.98$ performs quiet well on this metric.\\
 \indent To simplify the search space, we next fix our \textit{r-i} cut at the optimal value and analyze the FOM when varying the two parameters associated with the \textit{r-W1} cut. In Figure~\ref{fig:slope_int_cos}, we plot the FOM as a function of the slope and the intercept of the \textit{r-W1} threshold line. We overplot the contours of constant FOM to highlight the linear nature. It is clear that the FOM depends primarily on a fixed combination of these parameters. As a result, the optimal value of 
the slope of the line, $S_{r-W1}$, given a value the intercept, $i_{r-W1}$, can be obtained from the relation:
 \begin{eqnarray}   
      S_{r-W1}= 2.0 - 0.4 \times i_{r-W1}.\label{eqn:slope_int_eqn}
\end{eqnarray}
 Although the exact form of this optimal relation is dependent on the value of $b$ in Equation~\ref{eqn:fom_cos}, the existence of this correlation is significant, as it reduces our selection algorithm from a 3-parameter to a 2-parameter problem. We find that similar linear correlations occur for different values of $b$ in Equation~\ref{eqn:fom_cos} (e.g. $b\ =$ 0.25 or 0.5). In contrast to FOM, the normalized FOM depends little on the slope/intercept in the relevant parameter range, so in this case, our decisions are driven by FOM. \\
\indent Based on this approximation, we define a new variable $v$:  
 \begin{eqnarray}   
      v = (2.0 - S_{r-W1}) - 0.4 \times  i_{r-W1}.\label{eqn:v_eqn}
\end{eqnarray}
  This is defined such that $v$=0 at our nominal parameter values for the \textit{r-W1} cut. Next, we create a 2-dimensional grid to tabulate the FOM at each possible combination of the two parameters in our model, the minimum 
allowed \textit{r-i} and $v$. This grid is then analyzed to determine the parameter values which maximize the FOM. In Figure~\ref{fig:fom_ri}, we have plotted the maximum of the FOM over all $i_{r-W1}$ values as a function of the 
\textit{r-i} color cut. The maximum FOM decreases monotonically as the threshold \textit{r-i} color moves redward (corresponding to moving the vertical line in Figure~\ref{fig:rw1_cos} to the right). This result is consistent with our 
expectation, since we select a decreasing number of both LRGs and high-$z$ blue galaxies as the \textit{r-i} cut is moved redward.\\

\begin{figure}[h]
\centering
\includegraphics[scale=0.33]{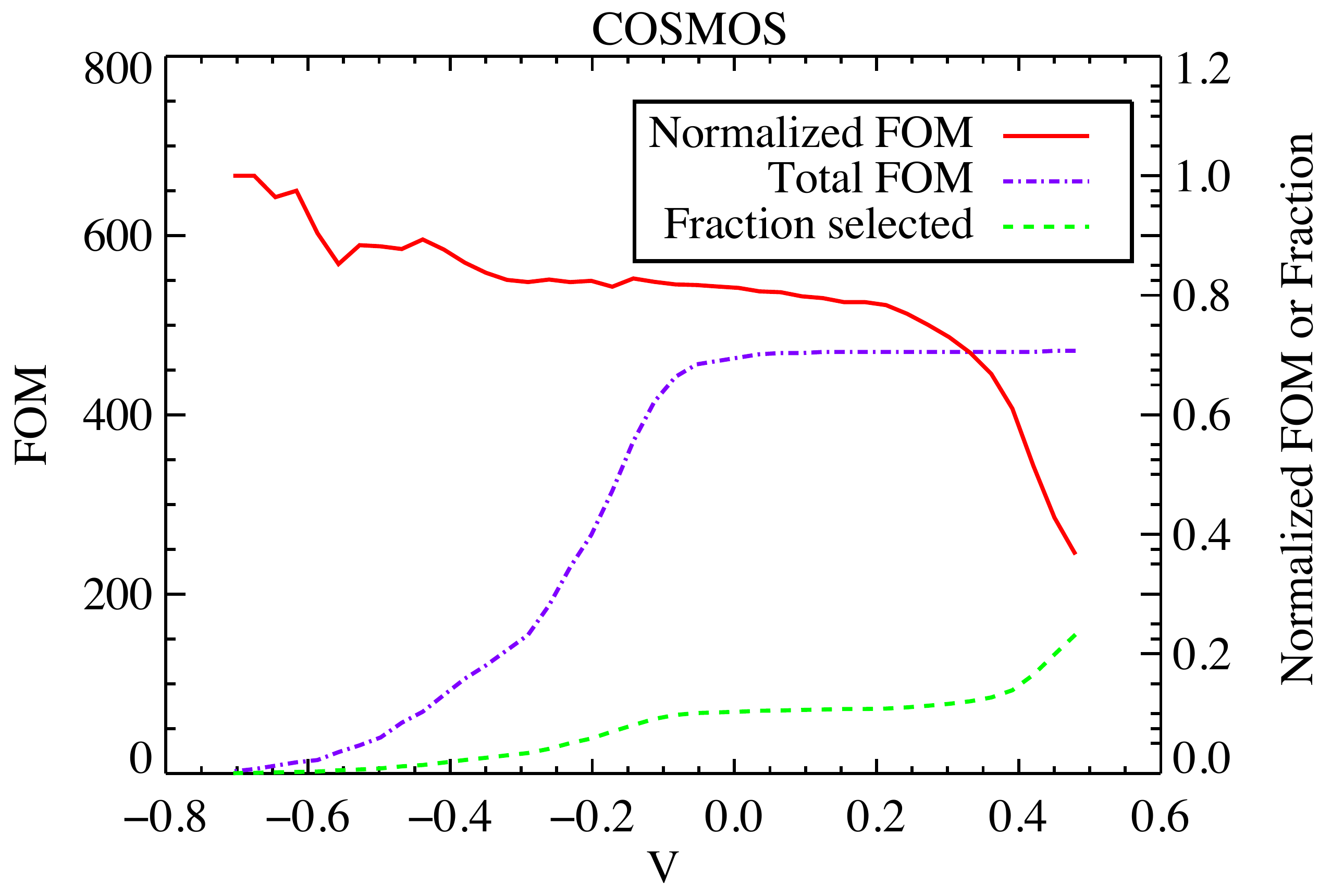}
\caption{\textsl{The FOM at \textit{r-i} $=$ 0.98 for z$ <$ 20.5 galaxies as a function of the combined slope/intercept parameter ($v$) in the COSMOS region. Only galaxies above a threshold redshift of $z >$ 0.6 are counted in the FOM. Each individual curve 
shows a different quantity. The second y-axis (on the right) is used for plotting both the
Normalized FOM (FOM normalized by the total number of objects selected
by a given color cut) and the fraction of all $z$ $<$ 20.5 objects that are
targeted. Values of $v$ around 0 are roughly optimal for both the total and normalized FOM.}}
\label{fig:fom_v_cos}
\end{figure}

\begin{figure}[h]
\centering
\includegraphics[scale=0.33]{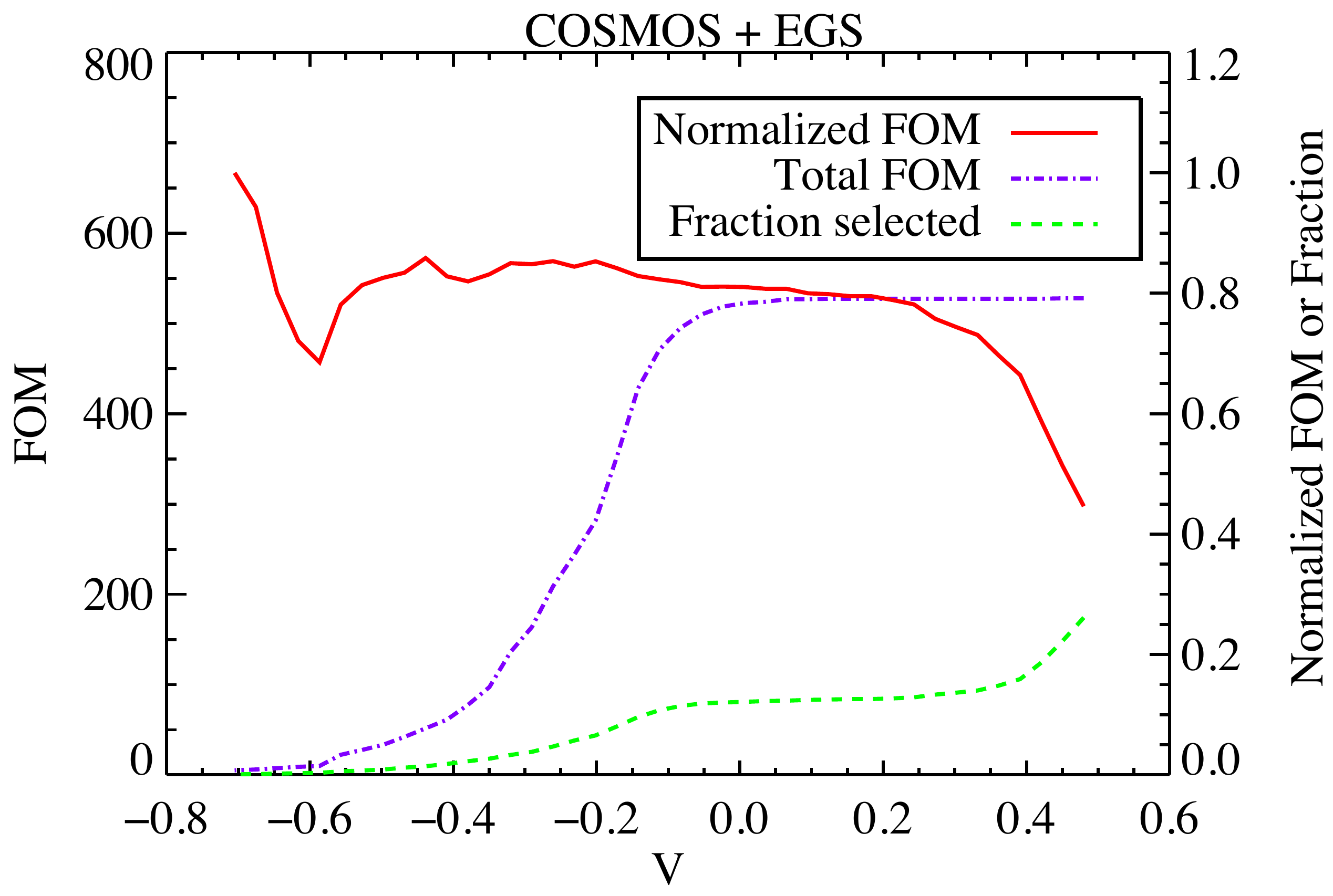}
\caption{\textsl{The FOM for \textit{r-i} $=$ 0.98 as a function of the combined slope/intercept parameter ($v$), averaged over both the COSMOS and the EGS fields. Only galaxies above a threshold redshift of $z >$ 0.6 are countedd in the FOM. Similar to Figure~\ref{fig:fom_v_cos}, values of $v$ around 0 prove to be roughly optimal for both the total and normalized FOM.}}
\label{fig:com_fom_v}
\end{figure}

 \indent In Figure~\ref{fig:fom_v_cos}, we have shown the FOM and normalized FOM as a function of $v$ assuming \textit{r-i} $=$ 0.98. We have also overplotted the fraction of objects selected as a consistency check. The flatness of the 
curves in Figure~\ref{fig:fom_v_cos} can be understood with the help of Figure~\ref{fig:rw1_cos}, as well as Equations~\ref{eqn:slope_int_eqn} and \ref{eqn:v_eqn}. As $v$ becomes positive, the intercept of the \textit{r-W1} 
threshold line becomes negative. This region lies within the empty region separating stars and galaxies in Figure~\ref{fig:rw1_cos}. Hence, we do not see any significant change in any of the quantities plotted. As $v$ is further 
increased, the \textit{r-W1} cut starts including stars into our sample. This causes an abrupt increase in the fraction of objects selected and a corresponding abrupt decrease in the FOM normalized by the number of selected 
objects. However, the total FOM remains mostly unaffected as stars do not contribute to it.\\
\indent Figure~\ref{fig:roc_cos} illustrates that \textit{r-i} $=$ 0.98 is a well-optimized cut for our purposes. Any decrease in this parameter causes the FPR to increase significantly without any significant 
gain in the TPR. Overall, we conclude that our nominal cuts of \textit{r-i} $> 0.98$ and \textit{r-W1} $> 2.0$ $\times$ \textit{(r-i)}, are well optimized and are the final cuts used in this selection. We find that when varying our threshold redshift 
marginally, e.g. $z$ $>$ 0.55 or $z$ $>$ 0.65, our analysis yields similar results for the optimal value of selection parameters.\\ 
\indent The analyses described so far all rely on our COSMOS field sample. It is worthwhile to test whether adding more data from different regions of the sky, which will reduce sample/cosmic 
variance, can improve our optimization. As explained in Section 2, in the EGS, we have obtained DEEP2 extended photometry from the catalog of \cite{Dan2013}. We repeat the same analysis as was done for the COSMOS field to estimate the FOM for our 2-parameter model. However, to estimate the total number of objects selected by a given color cut, we need to estimate the stellar contamination. Since DEEP2 avoided targeting stars \citep{jeff2013}, this is done separately using the HST-ACS general catalog of \citet{Griffith2012}, as described in Section 2.4. We otherwise repeat the same analysis as done for the COSMOS field to estimate the stellar contamination for a given color cut in 2D parameter space. The FOM in the EGS region shows a very similar behavior as in the COSMOS field as can be seen in Figure~\ref{fig:fom_ri} and \ref{fig:fom_v_cos}, indicating that sample/cosmic variance is not a major issue.\\
\indent In Figure~\ref{fig:com_fom_v}, we show the behavior of the FOM averaged over both the EGS + COSMOS fields. The plot shows a very consistent behavior, enabling us to conclude that our baseline color selection, \textit{r-i} $> 0.98$ and \textit{r-W1} $> 2.0$ $\times$ \textit{(r-i)} is indeed well-suited for selecting $z >$ 0.6 LRGs using this method.

\section{Conclusions and Future works} \label{sec:conclusions}
We have found a reliable and efficient method of identifying and selecting LRGs at higher redshifts by combining optical and infrared photometry. We have explored a variety of methods for optimizing our color cuts, given a particular set of rest-frame color and redshift requirements. With these optimization procedures, we can, for instance, tune the redshift range to select LRGs as required by different surveys.\\
\indent These methods have now been used to assemble large samples of LRGs. More than 10,000 \textit{z} $< $ 20, SDSS+WISE selected LRGs were targeted by a BOSS Ancillary program in 2012-2013 (SDSS DR12, in prep.). This will not only provide a good check on our selection methods but will also greatly increase the sample that we can use to optimize the selection process further. We have also selected LRGs based on similar methods, but using colors derived only from SDSS $ \textit{i}$,$ \textit{z}$ and \textit{WISE} $\textit{W1}$ (i.e., using \textit{i-W1} and \textit{i-z} colors). Selection in these redder bands helps for targeting higher-$z$ LRGs, but they are not as efficient as the combination of \textit{r}, \textit{i}, and \textit{W1} in star-galaxy separation.\\
\indent We have created a sample of LRGs over the entire SDSS footprint which has been used in selecting targets for a second BOSS ancillary survey, the SDSS Extended Quasars, Emission line galaxy, and LRGs Survey (SEQUELS). In SEQUELS, we have targeted $\sim 70{,}000$ \textit{z} $<$ 20 LRGs selected by one of two color cuts: One which utilizes \textit{r-i}, \textit{r-W1}, and a minimum value of \textit{i-z}, and a second which uses only \textit{i-z} and \textit{i-W1} colors. The work of analyzing the resulting spectra is in progress and will be reported in future publications. The same methods are being used for selecting LRG targets for eBOSS, which began observations in Fall 2014. The results from these selection algorithms will be described in future paper (Prakash et al. 2015).\\
\indent We are also investigating even deeper selections of LRGs using \textit{r}, \textit{z}, and \textit{W1} photometry for the proposed DESI survey (see DESI Conceptual Design Report).\footnote{\url{http://desi.lbl.gov/cdr/}} Optical/IR LRG selections have proved to be effective in our tests, and will provide a cornerstone sample for BAO surveys through the next decade.

\section*{Acknowledgements}
This work was supported by a U.S. Department of Energy Early Career grant. We thank Dan Matthews, David Schlegel, Arjun Dey, and Eduardo Rozo for helpful discussions.\\
\indent Funding for the DEEP2 survey has been provided by NSF grants AST-0071048, AST-0071198, AST-0507428, and AST-0507483. (Some of) The data presented herein were obtained at the W. M. Keck Observatory, which is operated as a scientific partnership among the California Institute of Technology, the University of California and the National Aeronautics and Space Administration. The Observatory was made possible by the generous financial support of the W. M. Keck Foundation. The DEEP2 team and Keck Observatory acknowledge the very significant cultural role and reverence that the summit of Mauna Kea has always had within the indigenous Hawaiian community and appreciate the opportunity to conduct observations from this mountain.\\
\indent This research has made use of the NASA/ IPAC Infrared Science Archive, which is operated by the Jet Propulsion Laboratory, California Institute of Technology, under contract with the National Aeronautics and Space Administration.\\
\indent This publication makes use of data products from the Wide-field Infrared Survey Explorer, which is a joint project of the University of California, Los Angeles, and the Jet Propulsion Laboratory/California Institute of Technology, funded by the National Aeronautics and Space Administration.\\
\indent CFHT LS is based on observations obtained with MegaPrime/MegaCam, a joint project of CFHT and CEA/IRFU, at the Canada-France-Hawaii Telescope (CFHT) which is operated by the National Research Council (NRC) of Canada, the Institut National des Science de l'Univers of the Centre National de la Recherche Scientifique (CNRS) of France, and the University of Hawaii. This work is based in part on data products produced at Terapix available at the Canadian Astronomy Data Centre as part of the Canada-France-Hawaii Telescope Legacy Survey, a collaborative project of NRC and CNRS.\\
\indent We gratefully acknowledge the contributions of the entire COSMOS collaboration consisting of more than 100 scientists. The HST COSMOS program was supported through NASA grant HST-GO-09822.

\clearpage

\bibliographystyle{apj} 
\bibliography{lrg_ref}\scriptsize

\end{document}